\documentclass[aps,showpacs,superscriptadress,preprint]{revtex4}
\usepackage{graphicx}
\usepackage{makecell}

\begin{document}
\title{The Big Bang Nucleosynthesis abundances of the light elements using improved thermonuclear reaction rates}
\author{ Chen Wu$^{1}$\footnote{Electronic address: wuchenoffd@gmail.com} }
\affiliation{
\small 1. Shanghai Advanced Research Institute, Chinese Academy of Sciences, Shanghai 201210, China}

\begin{abstract}
Big Bang Nucleosynthesis (BBN) is an important stage of a homogeneous and isotropic expanding universe.
 The results of calculation of the synthesis of light elements during this epoch can then be compared with the abundances of the light elements.
 The theoretical calculation of the  BBN model depends on the initial conditions  of the early  universe and  reaction cross sections measured by the nuclear physics experiment.   Recently, an update of the NACRE  (Nuclear Astrophysics Compilation of REactions)  database  is presented. This improved compilation comprises thermonuclear reaction rates for 34  two-body  reactions on light nuclides (fifteen are particle transfer reactions and  nineteen are radiative capture reactions).  In this work, we calculate the BBN abundances  by  using these updated thermonuclear reaction rates  in the framework of the  code \texttt{AlterBBN}.   Our results suggest that the new numerical result of the primordial Lithium abundance is 7.1 percent larger  than the previous calculation.
\end{abstract}

 \maketitle

BBN is one of the fundamental pillars of the cosmological standard model. Based on BBN theoretical model it is possible to calculate the abundances of the primordial light nuclides  via a network of nuclear processes, which can then be compared with the observed abundances. This allows to research the universe properties during BBN epoch, which is the most ancient period observationally accessible.   In standard cosmology,  the dynamics of this epoch is controlled by only one free parameter, the baryon to photon ratio, which can  be deduced from the observation of the Cosmic Microwave Background \cite{wmap}.

 To calculate the abundances of the light nuclides in the standard model of cosmology, several public codes, such as  PArthENoPE \cite{parthenope} or  \texttt{AlterBBN} \cite{alterBBN} already exist.  The code \texttt{AlterBBN} provides a fast and reliable calculation of the abundance of the elements in the standard model of cosmology as well as in alternative scenarios, which descends from the original code \texttt{NUC123} \cite{Kawano}. The code \texttt{AlterBBN} has been rewritten in C respecting the \texttt{C99} standard, and it has been applied in many theoretical investigation.

During the BBN stage, the Universe contains photons $\gamma$, leptons $e^-$,  $e^+$ and neutrinos $\nu$, hadrons $p$ and $n$.
 At the time of BBN, new nuclei will form over nuclear reactions (see Table I). In the following, we describe first the physics of BBN, and present the set of equations in the standard model of cosmology (we use the natural unit  $c = \hbar = k =1$). We then discuss the updated thermonuclear reaction rates and its effect on the calculated abundance of the light elements.

The cosmological expansion rate $\dot{a}$ governed by the Friedmann equation is  a function of the total density $\rho_{tot}$:
\begin{equation}
	H^2 \equiv \left( \frac{\dot{a}}{a} \right)^2 = \frac{8\pi G \rho_{tot}}{3}\,,
\end{equation}
where $H$ is defined as the Hubble parameter and $G$ is the gravitational constant. The equation of energy conservation is given by:
\begin{equation}
%%\frac{d}{dt}(\rho_{tot} a^3) + P_{tot} \frac{d}{dt}(a^3) - a^3 \left.\frac{d\rho_{tot}}{dt} = 0\;,
\dot{\rho}_{tot} = -3 H(\rho_{tot}+P_{tot}),
\end{equation}
in these equations, the total density $\rho_{tot}$ and pressure $P_{tot}$ are given by the sums over all the aforementioned constituents:
\begin{equation}
	\rho_{tot} = \rho_\gamma + \rho_\nu + \rho_{b} + \rho_{e^-} + \rho_{e^+} .
\end{equation}
\begin{equation}
P_{tot} = P_\gamma + P_\nu + P_b + P_{e^-}+P_{e^+}.
\end{equation}

for photons, The energy density  and pressure can be calculated through statistical mechanics:
\begin{equation}
\rho_\gamma=\frac{\pi^2 T^4}{15} \; , \qquad  P_\gamma = \frac{\pi^2 T^4}{45} ,
\end{equation}
and for neutrino:
\begin{equation}
\rho_\nu= N_\nu \frac{7}{8} \,\frac{\pi}{15} T_\nu^4 = N_\nu \frac{7\pi}{120} \left(\frac{4}{11}\right)^{4/3} T^4\; , \qquad  P_\nu = \frac{1}{3} \rho_\nu ,
\end{equation}
where $N_\nu=3$ is the number of neutrino families.  The factor $T_\nu/T =(4/11)^{1/3}$ comes from the interaction decoupling between hadrons and neutrino.

Then we write the sums of two leptons densities and pressures in terms of the modified Bessel functions $K_i$:
\begin{equation}
	\rho_{e^-} + \rho_{e^+} = \frac{2 m_e^4}{\pi^2} \sum_{n=1}^{\infty} (-1)^{n+1}\frac{1}{nz}\left( \frac{3}{4}K_3(nz) + \frac{1}{4}K_1(nz) \right)\cosh(n\phi_e)\,,
\end{equation}
\begin{equation}
	P_{e^-} + P_{e^+} = \frac{2 m_e^4}{\pi^2} \sum_{n=1}^{\infty} \frac{(-1)^{n+1}}{nz}\frac{K_2(nz)}{nz}\cosh(n\phi_e)\,,
\end{equation}
where we have defined the quantity $z = m_e/T$ and chemical potential $\phi_{e^-}  = \mu_e/T$.

The charge conservation of the Universe gives :
\begin{equation}
	n_{e^-} - n_{e^+} = \frac{h_\eta T^3 \sum_i Z_i Y_i}{M_{u}}\,,
\end{equation}
where $M_{u}$ is the unit atomic mass, $Z_i$ and $Y_i$ are the charge number and abundance of nuclei species $i$, respectively.

The baryon-to-photon ratio $\eta$ is determined  in the following way:
\begin{equation}
	h_\eta(T) = M_{u} \eta(T)\frac{n_\gamma(T)}{T^3}\,,
\end{equation}
where  h is parameter, $n_\gamma$ is the number density of photon. The difference in Eq. (9) can also be written as:
\begin{equation}
	n_{e^-} - n_{e^+} = \frac{2 m_e^3}{\pi^2} \sum_{i=1}^{\infty}(-1)^{n+1}\frac{K_2(nz)}{nz}\sinh(n\phi_e)\,.
\end{equation}
By using of the Eqs. (9) and (11), the electron chemical potential $\phi_e$ is  determined:
\begin{equation}
	\frac{d\phi_e}{dt} = \frac{\partial \phi_e}{\partial T}\frac{dT}{dt}  + \frac{\partial\phi_e}{\partial S}\frac{{d}S}{{d}t} + \frac{\partial \phi_e}{\partial a}\frac{{d}a}{{d}t}\,.
\end{equation}

On the other hand, the baryon energy density and pressure are determined as the sums on the $i$ nuclide, which can be found also in the manual of AlterBBN.
%\begin{equation}
%	\rho_{b} = \left( \sum_i\left( \frac{\Delta M_i}{M_{u}} + \zeta T \right)Y_i +1 \right)h_\eta T^3\,,
%\end{equation}
%\begin{equation}
%	P_{b} = \left( \frac{2}{3}\zeta T \sum_i Y_i \right)h_\eta T^3\,,
%\end{equation}
%where $\zeta = 1.388 \times 10 ^{-4}$  and  $\Delta M_i$ is the mass excess of nuclei $i$. The parameter $h_\eta$ can be determined by the relation:
%\begin{equation}
%	\frac{1}{h_\eta}\frac{{d}h_\eta}{{d}t} = -3 \frac{{d}\ln(T)}{{d}t} -3    \frac{{d}\ln(a)}{{d}t}\,.
%\end{equation}

The set of nuclear reactions used in  \verb?AlterBBN? can be written as  the  general form:
\begin{equation}
N_i\,^{A_i}\!Z_i \,+\, N_j\,^{A_j}\!Z_j \,\longleftrightarrow\, \, N_l\,^{A_l}\!Z_l  +  N_k\,^{A_k}\!Z_k \,\;.
\end{equation}
Therefore, the abundance evolution for nuclei $i$ is deduced by the equation \cite{Kawano}:
\begin{equation}
\frac{dY_i}{dt}=\sum_{j,k,l} N_i \left(-\frac{Y_i^{N_i} Y_j^{N_j}}{N_i!\, N_j!} \Gamma_{ij}^k + \frac{Y_l^{N_l} Y_k^{N_k}}{N_l!\, N_k!} \Gamma_{lk}^j \right) \;,
\end{equation}
where the nuclide abundance  $Y_i$ is $ X_i / A_i$, $X_i$ is the mass fraction in nuclide $i$ and $A_i$ is their atomic number. $N_i$ is the number of nuclides $i$ that enters into the nuclear reaction,
 $\Gamma_{ij}^k$ and $\Gamma_{lk}^j$  are the forward and reverse reaction rates respectively.  \verb?AlterBBN?  includes a network of 88 nuclear reactions, which are gathered in the Table I. The code sets the number of neutrino species to  3.0, the baryon-to-photon ratio to $\eta_0$ $= 6.09\times 10^{-10}$ \cite{Planck} , the initial temperature to  $2.7\times 10^{10}\,$K (corresponding to $2.3\,$MeV) and the lifetime of the neutron to  879.4 seconds \cite{neutron}. Given proper initial conditions, the set of Eqs. (1), (2), (10), (12) and (14) will be solved by a 4th-order Runge-Kutta integration.

Observational measurements  are cited from the latest report:
\begin{equation}
\textmd{Y}_{p} = 0.2453 \pm 0.0034 \,,
\end{equation}
\begin{equation}
^3 \textmd{He}/\textmd{H} < (1.1 \pm 0.2) \times 10^{-5}\,,
\end{equation}
\begin{equation}
^2 \textmd{H/H} = (2.527  \pm 0.030) \times 10^{-5}\,,
\end{equation}
\begin{equation}
^7 \textmd{Li}/\textmd{H} = (1.58^{+0.35}_{-0.28}) \times 10^{-10} \,.
\end{equation}
which constrains the helium abundance $\textmd{Y}_p$ \cite{He4} and the primordial $^2 \textmd{H/H}$ \cite{H2}, $^3 \textmd{He}/\textmd{H}$ \cite{He3} and $^7 \textmd{Li}/\textmd{H}$ \cite{Li7} ratios.

On the other hand, the thermonuclear reaction rates in  nuclear reaction network  play an significant role
in big-bang nucleosynthesis investigations. The thermonuclear reaction rates of a two-body reaction $A(a,b)B$ are  determined as the Avogadro number $N_{\rm A}$ times  the Maxwellian-averaged rate $<\sigma v>$,
\begin{equation}
 <\sigma v> =  \frac{(8/\pi)^{1/2}}{\mu^{1/2} (k_{\rm B}T)^{3/2}} \int_{0}^{\infty} E\, \sigma(E)\, {\rm exp}[- E/(k_{\rm B}T)]\, {\rm d}E,
\label{eqrate}
\end{equation}
\noindent
where $v$ being the relative velocity between the target ($A$) and projectile ($a$) nuclei, $\sigma(E)$ is the reaction cross section at the center-of-mass incident energy $E $, $\mu $ is the reduced mass,  $T$ is the temperature and  $k_{B}$ is the Boltzmann constant.

 The series of advance of William A. Fowler and his coworkers \cite{FCZ67, CF88}  are pioneering work in this research field. Then, the  so-called NACRE (Nuclear Astrophysics Compilation of REactions) database \cite{NACRE1.0} is a second important progress of such astrophysics-oriented compilations, which  comprise  an ensemble of 86 charged-particle induced reactions  involved in BBN. Since  1999, the NACRE database has indeed been used in many stellar  evolution models as well as of nucleosynthesis investigations.

  Since NACRE, many cross sections measure of astrophysical interest have been carried out, and many theoretical efforts have been investigated for better predictions of the concerned reaction rates.  For instance, thermonuclear reaction rates of relevance to the BBN have been recomputed by means  of the R-matrix method \cite{Rmatrix}.

  In parallel to these developments,  an update and an extension of NACRE is published in Ref. \cite{NPA2013}.   The new version of NACRE II comprises 34 two-body reactions (15 particle transfer and 19 radiative capture reactions), and uses nuclear potential models to phenomenologically  extrapolate resonant and nonresonant reaction cross sections to low energies of interest.   NACRE II featured (1) detailed references to the works of the experimental data (and to some nuclear theoretical works); (2) the extrapolation of astrophysical $S$-factors to very low energies based on potential models; (3)  a tabular presentation of the nuclear reaction rates in the $10^{6} < T \leq 10^{10}K$  temperature range.  (See \cite{NPA2013} for more details).

In Table I of  Ref. \cite{NPA2013}, the results for the 34 two-body exoergic reactions  are summarized, of which 29 two-body reactions are implemented in the
\texttt{AlterBBN} network of nuclear reactions.    Specifically, the 29 nuclear reactions are the following nuclear reactions in Table I:
 No. 19, 21, 22, 23, 24, 25, 26, 27, 28, 29, 30, 31, 39, 40, 41 ,42,  45, 46,  49, 50, 53, 71, 72, 74, 75, 76, 77, 78, 87.

\begin{tabular}
{p{4.0cm}p{4.2cm}p{4.2cm}p{4.2cm}}
\multicolumn{4}{c}{TABLE II. The abundances of the light elements using improved thermonuclear reaction rates  }\\
\hline
   &  Observations      &    A. Arbey (2012) \cite{alterBBN}      &  this work       \\
\hline
 $Y_p$                                    & 0.2453 $\pm 0.0034$ \cite{He4}		& 0.2476	&  0.2462 \\
  $^2$H/H ($\times 10^{-5}$)            & 2.527	$\pm 0.030$  \cite{H2}	& 2.515	&  2.589 \\
$^3$He/H ($\times 10^{-5}$)            & $<1.1\pm0.2$ \cite{He3}	& 1.015	&  1.045 \\
 $^7$Li/H ($\times 10^{-10}$)           & $1.58^{+0.35}_{-0.28}$ \cite{Li7}		& 4.694	&  5.028 \\
 \hline
\end{tabular}
\\
\newline
In this work, we apply these  new thermonuclear reaction rates to calculate primordial abundances of the light elements generated   by using the code AlterBBN.
Our results are listed in Table II. The new calculated abundances of $^4\textmd{He}$, $^2\textmd{H}$ and $^3\textmd{He}$ nuclides are almost the same  compared with the old calculated  results. However, compared to the previous numerical results, there is about 7.1 percent increase in the abundances of the lithium nuclides.
 Recent observational measurement of $^7 \textmd{Li}$ abundance is $1.58^{+0.35}_{-0.28} \times10^{-10}$ \cite{Li7}, which is  almost three times smaller than that predicted by the BBN model. After using the updated thermonuclear reaction rates in the AlterBBN code, the calculated primordial abundance of  $^7 \textmd{Li}$  is $5.028\times 10^{-10}$, which make the primordial Lithium problem much worse.  Similar conclusions have been also observed by other work in literature \cite{Li7jcap}.

 Finally, we give some discussions and perspectives of this work.  Our work presents calculations of the abundances of the light elements produced in BBN. This is an active domain that aims at better precision, in order to match the high precision reached by observations. However, we must emphasize that some of the nuclear reaction rates used in our network calculations are referred to publications before 1994, as shown in the Table 1. In fact, the  nuclear reaction rates have been continually updated, more new nuclear reaction rates are reported in the Table 4 in literature  \cite{rates}. Moreover, many important reactions of interest for  the physics of BBN has been measured  recently.  For instance, One of the most relevant reactions for the $^7\textmd{Li}$ abundance is the $^7\textmd{Be}(n, \alpha)^4\textmd{He}$ reaction, so the $^7\textmd{Be}(n, \alpha)^4\textmd{He}$  nuclear reaction cross section has been measured for the first time from 10 meV to 10 keV neutron energy \cite{prl}. In this report, the authors show that their  results hint to a minor role of this reaction in BBN, leaving the long-standing primordial Lithium problem unsolved. Since new experimental data have become available,  we  would try to include  the progress of these experiments in future theoretical calculations.

\begin{table}[ht]
\hspace*{-1.cm}\mbox{\begin{tabular}{|c|c|c|}
\hline
 No.  & Reaction  & Ref.\\
\hline\hline
0  & $n \leftrightarrow p + e^- + \bar\nu$   & \cite{smith92}\\
\hline\hline
1  & $^3\!H \rightarrow e^- + \nu + ^3\!He$ & \cite{tilly87}\\
\hline
2  & $^8\!Li \rightarrow e^- + \nu + 2\,^4\!He$ & \cite{ajzenberg88}\\
\hline
3  & $^{12}\!B \rightarrow e^- + \nu + ^{12}\!C$ & \cite{ajzenberg90}\\
\hline
4  & $^{14}\!C \rightarrow e^- + \nu + ^{14}\!N$  & \cite{ajzenberg86}\\
\hline
5  & $^8\!B \rightarrow e^+ + \nu + 2\,^4\!He$  & \cite{ajzenberg88}\\
\hline
6 & $^{11}\!C \rightarrow e^+ + \nu + ^{11}\!B$   & \cite{ajzenberg90}\\
\hline
7  & $^{12}\!N \rightarrow e^+ + \nu + ^{12}\!C$  & \cite{ajzenberg90}\\
\hline
8  & $^{13}\!N \rightarrow e^+ + \nu + ^{13}\!C$  & \cite{ajzenberg86} \\
\hline
9  & $^{14}\!O \rightarrow e^+ + \nu + ^{14}\!N$  & \cite{ajzenberg86} \\
\hline
10  & $^{15}\!O \rightarrow e^+ + \nu + ^{15}\!N$  & \cite{ajzenberg86} \\
\hline\hline
11  & $H + n \rightarrow \gamma + ^2\!H$  & \cite{smith92} \\
\hline
12  & $^2\!H + n \rightarrow \gamma + ^3\!H$  & \cite{wagoner69} \\
\hline
13  & $^3\!He + n \rightarrow \gamma + ^4\!He$ & \cite{wagoner69} \\
\hline
14  & $^6\!Li + n \rightarrow \gamma + ^7\!Li$  & \cite{malaney93} \\
\hline
15  & $^3\!He + n \rightarrow p + ^3\!H$  & \cite{smith92} \\
\hline
16  & $^7\!Be + n \rightarrow p + ^7\!Li$  & \cite{smith92} \\
\hline
17  & $^6\!Li + n \rightarrow \alpha + ^3\!H$  & \cite{caughlan88} \\
\hline
18  & $^7\!Be + n \rightarrow \alpha + ^4\!He$  & \cite{wagoner69} \\
\hline
\underline{19} & $^2\!H + p \rightarrow \gamma + ^3\!He$   & \cite{smith92} \\
\hline
20  & $^3\!H + p \rightarrow \gamma + ^4\!He$ & \cite{smith92} \\
\hline
\underline{21}  & $^6\!Li + p \rightarrow \gamma + ^7\!Be$ & \cite{caughlan88} \\
\hline
\underline{22}  & $^6\!Li + p \rightarrow \alpha + ^3\!He$ & \cite{caughlan88} \\
\hline
\underline{23}  & $^7\!Li + p \rightarrow \alpha + ^4\!He$ & \cite{smith92} \\
\hline
\underline{24}  & $^2\!H + \alpha \rightarrow p + ^6\!Li$ & \cite{caughlan88} \\
\hline
\underline{25}  & $^3\!H + \alpha \rightarrow p + ^7\!Li$ & \cite{smith92} \\
\hline
\underline{26}  & $^3\!He + \alpha \rightarrow p + ^7\!Be$ & \cite{smith92} \\
\hline
\underline{27}  & $^2\!H  + D \rightarrow p + ^3\!He$ & \cite{smith92} \\
\hline
\underline{28}  & $^2\!H  + D \rightarrow n + ^3\!H$ & \cite{smith92} \\
\hline
\underline{29}  & $^3\!H  + D \rightarrow n + ^4\!He$ & \cite{smith92} \\
\hline
 \end{tabular}
\hspace*{-0.cm}\begin{tabular}{|c|c|c|}
\hline
  No.  & Reaction & Ref. \\
\hline\hline
\underline{30}  & $^3\!He  + D \rightarrow p + ^4\!He$  & \cite{caughlan88} \\
\hline
\underline{31}  & $^3\!He + ^3\!He \rightarrow 2\,p + ^4\!He$ & \cite{caughlan88} \\
\hline
32  & $^7\!Li  + D \rightarrow n + \alpha + ^4\!He$ & \cite{caughlan88} \\
\hline
33 &   $^7\!Be  + D \rightarrow p + \alpha + ^4\!He$ &\cite{caughlan88} \\
\hline
34  & $^7\!Li + n \rightarrow \gamma + ^8\!Li$  & \cite{wagoner69} \\
\hline
35 & $^{10}\!B + n \rightarrow \gamma + ^{11}\!B$  & \cite{wagoner69} \\
\hline
36  & $^{11}\!B + n \rightarrow \gamma + ^{12}\!B$ & \cite{malaney93} \\
\hline
37  & $^{11}\!C + n \rightarrow p + ^{11}\!B$ & \cite{caughlan88} \\
\hline
38  & $^{10}\!B + n \rightarrow \alpha + ^7\!Li$ & \cite{caughlan88} \\
\hline
\underline{39}  & $^7\!Be + p \rightarrow \gamma + ^8\!B$  & \cite{caughlan88} \\
\hline
\underline{40}  & $^9\!Be + p \rightarrow \gamma + ^{10}\!B$  & \cite{caughlan88} \\
\hline
\underline{41}  & $^{10}\!B + p \rightarrow \gamma + ^{11}\!C$ & \cite{caughlan88} \\
\hline
\underline{42}  & $^{11}\!B + p \rightarrow \gamma + ^{12}\!C$ & \cite{caughlan88} \\
\hline
43  & $^{11}\!C + p \rightarrow \gamma + ^{12}\!N$ & \cite{caughlan88} \\
\hline
44  & $^{12}\!B + p \rightarrow n + ^{12}\!C$ & \cite{wagoner69} \\
\hline
\underline{45}  & $^9\!Be + p \rightarrow \alpha + ^6\!Li$ & \cite{caughlan88} \\
\hline
\underline{46}  & $^{10}\!B + p \rightarrow \alpha + ^7\!Be$ & \cite{caughlan88} \\
\hline
47  & $^{12}\!B + p \rightarrow \alpha + ^9\!Be$ & \cite{wagoner69} \\
\hline
48  & $^6\!Li + \alpha \rightarrow \gamma + ^{10}\!B$ & \cite{caughlan88} \\
\hline
\underline{49}  & $^7\!Li + \alpha \rightarrow \gamma + ^{11}\!B$ & \cite{caughlan88} \\
\hline
\underline{50}  & $^7\!Be + \alpha \rightarrow \gamma + ^{11}\!C$ & \cite{caughlan88} \\
\hline
51  & $^8\!B + \alpha \rightarrow p + ^{11}\!C$ & \cite{wagoner69} \\
\hline
52  & $^8\!Li + \alpha \rightarrow n + ^{11}\!B$ & \cite{malaney93} \\
\hline
\underline{53}  & $^9\!Be + \alpha \rightarrow n + ^{12}\!C$ & \cite{caughlan88} \\
\hline
54  & $^9\!Be  + D \rightarrow n + ^{10}\!B$ & \cite{Kawano} \\
\hline
55  & $^{10}\!B  + D \rightarrow p + ^{11}\!B$ & \cite{Kawano} \\
\hline
56 & $^{11}\!B  + D \rightarrow n + ^{12}\!C$ & \cite{Kawano}  \\
\hline
57  & $^4\!He + \alpha + n \rightarrow \gamma + ^9\!Be$ & \cite{caughlan88} \\
\hline
58  & $^4\!He + 2\,\alpha \rightarrow \gamma + ^{12}\!C$ & \cite{caughlan88} \\
\hline
59  & $^8\!Li + p \rightarrow n + \alpha + ^4\!He$  & \cite{Kawano}  \\
\hline
 \end{tabular}
\hspace*{-0.cm}\begin{tabular}{|c|c|c|}
\hline
  No.  & Reaction & Ref. \\
\hline\hline
60  & $^8\!B + n \rightarrow p + \alpha + ^4\!He$ & \cite{Kawano} \\
\hline
61  & $^9\!Be + p \rightarrow d + \alpha + ^4\!He$ & \cite{caughlan88} \\
\hline
62  & $^{11}\!B + p \rightarrow 2\,\alpha + Be4$ & \cite{caughlan88} \\
\hline
63  & $^{11}\!C + n \rightarrow 2\,\alpha + ^4\!He$ & \cite{wagoner69} \\
\hline
64  & $^{12}\!C + n \rightarrow \gamma + ^{13}\!C$ & \cite{wagoner69} \\
\hline
65  & $^{13}\!C + n \rightarrow \gamma + ^{14}\!C$ & \cite{wagoner69} \\
\hline
66  & $^{14}\!N + n \rightarrow \gamma + ^{15}\!N$ & \cite{wagoner69} \\
\hline
67  & $^{13}\!N + n \rightarrow p + ^{13}\!C$ & \cite{caughlan88} \\
\hline
68  & $^{14}\!N + n \rightarrow p + ^{14}\!C$ & \cite{caughlan88} \\
\hline
69  & $^{15}\!O + n \rightarrow p + ^{15}\!N$ & \cite{caughlan88} \\
\hline
70  & $^{15}\!O + n \rightarrow \alpha + ^{12}\!C$ & \cite{caughlan88} \\
\hline
\underline{71}  & $^{12}\!C + p \rightarrow \gamma + ^{13}\!N$ & \cite{caughlan88} \\
\hline
\underline{72}  & $^{13}\!C + p \rightarrow \gamma + ^{14}\!N$ & \cite{caughlan88} \\
\hline
73  & $^{14}\!C + p \rightarrow \gamma + ^{15}\!N$ & \cite{caughlan88} \\
\hline
\underline{74}  & $^{13}\!N + p \rightarrow \gamma + O14$ & \cite{caughlan88} \\
\hline
\underline{75}  & $^{14}\!N + p \rightarrow \gamma + ^{15}\!O$ & \cite{caughlan88} \\
\hline
\underline{76}  & $^{15}\!N + p \rightarrow \gamma + ^{16}\!O$ & \cite{caughlan88} \\
\hline
\underline{77}  & $^{15}\!N + p \rightarrow \alpha + ^{12}\!C$ & \cite{caughlan88} \\
\hline
\underline{78}  & $^{12}\!C + \alpha \rightarrow \gamma + ^{16}\!O$ & \cite{caughlan88} \\
\hline
79  & $^{10}\!B + \alpha \rightarrow p + ^{13}\!C$ & \cite{wagoner69} \\
\hline
80  & $^{11}\!B + \alpha \rightarrow p + ^{14}\!C$ & \cite{caughlan88} \\
\hline
81  & $^{11}\!C + \alpha \rightarrow p + ^{14}\!N$ & \cite{caughlan88} \\
\hline
82  & $^{12}\!N + \alpha \rightarrow p + ^{15}\!O$ & \cite{caughlan88} \\
\hline
83  & $^{13}\!N + \alpha \rightarrow p + ^{16}\!O$ & \cite{caughlan88}\\
\hline
84  & $^{10}\!B + \alpha \rightarrow n + ^{13}\!N$ & \cite{caughlan88} \\
\hline
85  & $^{11}\!B + \alpha \rightarrow n + ^{14}\!N$ & \cite{caughlan88} \\
\hline
86  & $^{12}\!B + \alpha \rightarrow n + ^{15}\!N$ & \cite{wagoner69} \\
\hline
\underline{87}  & $^{13}\!C + \alpha \rightarrow n + ^{16}\!O$ & \cite{caughlan88} \\
\hline
&&\\[0.65cm]
\hline
\end{tabular}}
\caption{Network of nuclear reactions implemented in {\tt AlterBBN}. If the serial number of the nuclear reaction is underlined, it indicates that the nuclear reaction rate is updated with new data \cite{NPA2013}.}
\end{table}

\begin{acknowledgments}
The authors thank Prof. Jun Xu, Fei Lu and Guoqiang Zhang for their positive help and useful discussion. We also like to thank developers of the AlterBBN packages for their opening codes.  This work was supported by the National Key Research and Develop Program of China under Contract No. 2018YFA0404404 and No. 2016YFA0400502.
\end{acknowledgments}

%\bibliography{NDSch-Ref}
\end{document}